\documentclass[a4paper,10pt]{report}
\usepackage[pdftex]{graphicx}
\usepackage{pgfplots}
\usepackage{geometry}
\usepackage{cite}
\usepackage{subfig}
\usepackage{amssymb}
\usepackage{amsmath}
\usepackage{tikz} % for diagram
\usepackage{multirow} %for the table
\usepackage{microtype}
\usepackage{xcolor}
\usepackage{chngcntr}% this package is for not star section with o.xx
\usepackage[breaklinks=true,
pdfborder={0 0 0},
bookmarks=true,
pdfhighlight=/O]{hyperref}

\date{}

\usetikzlibrary{arrows,shapes,positioning,shadows,trees}
\tikzset{
  basic/.style  = {draw, text width=2cm, drop shadow, font=\sffamily, rectangle},
  root/.style   = {basic, rounded corners=2pt, thin, align=center,
                   fill=black!20},
  level 2/.style = {basic, rounded corners=6pt, thin,align=center, fill={rgb:cyan,2;yellow,2;blue,1},
                   text width=8em},
  level 3/.style = {basic, thin, align=left, fill={rgb:cyan,3;yellow,1;blue,1}, text width=6.5em}
}

\title{Memristor nanodevice for unconventional computing: review and applications }

\author{
Mahyar Shahsavari, Pierre Boulet\\
 Univ. Lille, CNRS, Centrale Lille, UMR 9189 - CRIStAL \\ Centre de Recherche
 en Informatique Signal et Automatique de Lille,\\ F-59000 Lille, France.\protect\\
}
\begin{document}
%\pagestyle{headings}
%\setcounter{page}{1}
%\pagenumbering{roman}
\counterwithout{section}{chapter}% this command  is for not star section with o.xx

\maketitle
\begin{abstract}
A memristor is a two-terminal nanodevice that its properties 
attract a wide community of researchers from various domains such as
physics, chemistry, electronics, computer and 
neuroscience. The simple structure for manufacturing, small scalability, 
nonvolatility and potential of using in low power platforms are outstanding
characteristics of this emerging nanodevice. 
In this report, we review a brief literature of memristor from 
mathematic model to the physical realization.
We discuss different classes of memristors based on the material used for 
its manufacturing. The potential applications of memristor are
presented and a wide domain of applications are explained and classified. 

\end{abstract}
\tableofcontents
\newpage
\section{Introduction}
Memristor has recently drawn the wide attention of scientists and researchers due to non-volatility, better alignment, 
and excellent scalability properties \cite{LiH10}. Memristor has initiated a novel research direction
for the advancement of neuromorphic and neuro-inspired computing.
Memristor remembers its last state after the last power
plugging and has a simple physical structure, high-density integration, and 
low-power consumption. These features make the memristor an attractive candidate
for building the next generation of memories \cite{Ho:2009}. In addition,  
from high-performance computing point of view, the memristor has the potential capability to
conquer the memory bottleneck issue, by utilizing computational unit next to the memory \cite{Di-Ventra}.
Due to these unique properties and potentials of the memristor, neuroscientists and neuromorphic researchers apply 
it as an artificial synapse in Spiking Neural Network (SNN) architectures \cite{jo_nanoscale_2010}. 

Memristor was predicted in 1971 
by Leon Chua, a professor of electrical 
engineering at the University of California, Berkeley, as 
the fourth fundamental device  \cite{chua_memristor-missing_1971}. Publishing a paper in the Nature journal
by Hewlett Packard (HP) \cite{strukov_missing_2008} in May 2008, announced the first ever experimental
realization of
the memristor, caused an extraordinary increased interest in this passive
element. Based on the symmetry of the equations 
 that govern the resistor, capacitor, and inductor, Chua hypothesized that fourth device should exist
that holds a relationship between magnetic flux and charge. 
 After the physical discovery of the memristor, several institutions have published
the memristor device fabrications using a variety of different materials and device 
structures \cite{strukov_missing_2008,ferro,Berzina-poly,spin1,alibart_organic_2010}.  

In 2009, Biolek \textit{et al}. modeled nonlinear dopant drift memristor by SPICE \cite{Biolek2009}.
One year later, Wei Lu, professor at the University of Michigan proposed a nanoscale memristor device which can 
mimic the synapse behavior in 
neuromorphic systems \cite{Wei-syn-mem}. Later on, in 2011 a team of multidisciplinary researchers from Harvard 
University published an interesting paper on
programmable nanowire circuits for using in nanoprocessors \cite{Yan2011240}. Until June 2016, based on 
the Scopus bibliographic database, 2466 papers have been published in peer-reviewed journals and ISI articles 
which are related to memristor fabrication or applications of the memristor in different Domains of 
science and technology.

Memristors are promising devices for a wide range of
potential applications from digital memory, logic/analog
circuits, and bio-inspired applications \cite{rose10}.
Especially because the nonvolatility property in many types of memristors,they 
could be a suitable candidate for
making non-volatile memories with ultra large capacity \cite{Borghetti}. 
In addition to non-volatility, the memristor has other attractive features such as simple physical structure, 
high-density, low-power,
and unlimited endurance which make this device a proper choice for many applications.
Different device structures are still being developed to determine which
memristor device can be presented as the best choice for commercial use in memory/flash manufacturing 
or in neuromorphic platforms.
This is based on different
factors such as size, switching speed, power consumption, switching longevity, and
CMOS compatibility. The rest of the manuscript is organized as follows: In Section 2, a general overview of 
the memristor is done and the electrical 
properties have been investigated. 
Section 3 presents memristor implementation and fabrication. We investigate various types of
memristors based on the different materials that have been used in the fabrication. 
In Section 4, potential applications of Memristor has been studied. Section 5 deals with streams of research, 
we have investigated a research classification from the physics level to the system design.
Finally, we describe a brief summary and the future work. 
\section{Memristor device overview and properties}

In this section, we discuss the memristor nanodevice which is believed to be the fourth missing fundamental circuit
element, that comes in the form of a passive two-terminal device. We discuss how it can remember its state, 
and what is 
its electrical model and particular properties.
\subsection{Memristor a missing electrical passive element}
\begin{figure}
\centering
\includegraphics[width=7cm]{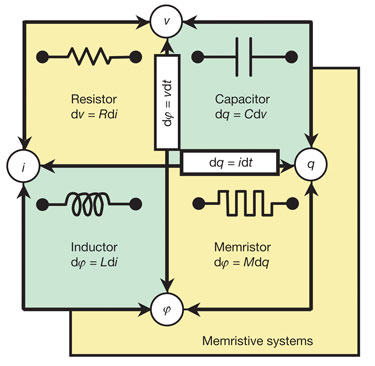}
\caption{Relations between the passive devices 
and the anticipating the place of the fourth fundamental element based on the
relations between charge ($q$) and flux ($\varphi)$ (from \cite{strukov_missing_2008}).}
\label{Memristor02}
\end{figure}

Memristor is a contraction of ``\textit{memory} \& \textit{resistor,}'' because the basic functionality of the
memristor is to remember its state history. This characteristic proposes a promising component for
next generation memory.
Memristor is a thin-film electrical circuit element that changes its
resistance depending on the total amount of charge that flows through the device.
Chua proved that memristor behavior could not be duplicated by any circuit built 
using only the other three basic electronic elements (Resistor,Capacitor, Inductor), 
that is why the memristor is truly fundamental. As it is depicted in Figure \ref{Memristor02},
the resistor is constant factor between the voltage and current
($dv=R.di$), the capacitor is a constant factor between the charge and voltage $(dq=C.dv)$,
and the inductor is a constant factor between the flux and current $(d\varphi=L.di)$.
The relation between flux and charge is Obviously missing $(d\varphi=M.dq)$ that 
can be interpreted by a fourth fundamental element such as memristor \cite{chua_memristor-missing_1971}.

Obviously, in memristive devices, the
nonlinear resistance can be changed and memorized by controlling the flow of
the electrical charge or the magnetic flux. This control any two-terminal black box is called a
memristor if, and only if, it exhibits a pinched hysteresis
loop for all bipolar periodic input current signaling is interesting 
for the computation capability of a device similar to the controlling of the states of a transistor. 
For instance in an analog domain, one can control the state of a transistor to stay in 
an active area for amplification. Nevertheless, in the digital domain to stay in Off (cut-off) state for logic '0'
and in On (saturated) state for logic '1' one can perform with controlling the gate voltage. 
The output current in MOSFET (Metal-Oxide semiconductor Field Effect Transistor) is managed  by changing
the gate voltage as well as in BJT (Bipolar Junction Transistor) the input current (base current)
can control the output current (collector-emitter current). The main difference between 
the memristor and transistor
for managing the states is that in transistor there is a third terminal to control the states however,
in contrast a memristor is a two-terminal device
and there is no extra terminal to control the device state.
The challenge of using memristor 
as a computational component instead of transistor lies in the ability to control the working states 
as accurate as possible. Indeed, in a memristor both potentials for analog and digital computing have 
been presented. Consequently, using memristor in digital computing
to make gate library or crossbar architecture as well as using memristor
in analog domain (neuro-inspired or traditional)
for computation are introduced in several work
\cite{Di-Ventra,Leveraging,LehtonenPL12,Crossbar-mem,indiveri_integration_2013-1}.
In next sections, we discuss different possibilities and our contributions to apply memristor in 
both digital and analog platforms.

\subsection{Memristive device functionality}
\label{howworks}
When you turn off the voltage, the memristor remembers its most recent resistance until the next time you 
turn it on, whether that happens a day later or a year later. It is worth mentioning that the duration  
to store the data in resistive form is dependent of the nano-device material. In other words, the 
volatility is different depending on the device materials in fabrication. 

To understand the functionality of a memristor, let us imagine a resistor as a pipe which water
flows through it. The water simulates the electric charge. 
The resistor obstruction of the flow of charge is comparable
to the diameter of the pipe: the narrower the pipe, the greater the resistance. For the history of 
circuit design, resistors have had a fixed pipe diameter.
But a memristor is a pipe that its  diameter changes with the amount and direction 
of the water flows through it. 
If water flows through this
pipe in one direction, it expands the pipe diameter (more conductive). 
But if the water flows in the opposite direction and the pipe shrinks (less conductive). Furthermore, 
let us imagine while we turn off the flow, the diameter of the pipe freezes 
until the water is turned back on.
It mimics the memristor characteristic to remember last state.
This freezing property suits memristors brilliantly for the new generation of memory. 
The ability to indefinitely store resistance values 
means that a memristor can be used as a nonvolatile memory.

\begin{figure}[ht!]
\centering
\includegraphics[height=5cm,width=8cm]{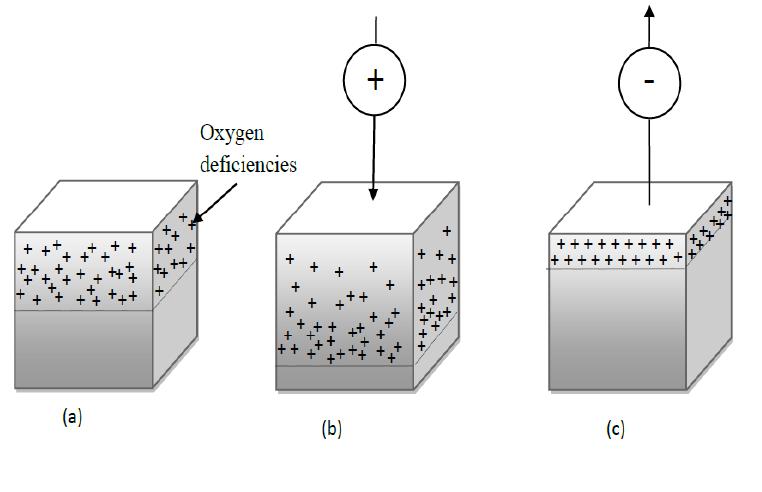}
\caption{A material model of the memristor schematic to
demonstrate TiO$_{2}$ memristor functionality, positive charge makes 
the device more conductive and negative charge makes it less conductive.
}
\label{fig:Dope}
\end{figure}

Chua demonstrated mathematically that his hypothetical device would provide a relationship between flux 
and charge similar to what a resistor provides between voltage and current.
There was no obvious physical interaction between charge and the integral over the voltage before HP discovery.  
Stanley Williams in \cite{Williams-missingmem} explained how they found the missing 
memristor and what is the relation between what they found and Chua mathematic model. 
In Figure \ref{fig:Dope}, the oxygen deficiencies in the TiO$_{2-x}$ manifest as bubbles
of oxygen vacancies scattered throughout the
upper layer. A positive voltage on the switch
repels the (positive) oxygen deficiencies in the
metallic upper TiO$_{2-x}$ layer, sending them into
the insulating TiO$_{2}$ layer below. That causes the
boundary between the two materials to move
down, increasing the percentage of conducting
TiO$_{2-x}$ and thus the conductivity of the entire
switch. Therefore, the more positive voltage causes the
more conductivity in the cube.
A negative voltage on the switch attracts the positively
charged oxygen bubbles, pulling them out
of the TiO$_{2}$. The amount of insulating of
resistive TiO$_{2}$ increases, thereby making the switch more resistive. 
The more negative voltage causes the less conductivity in the cube.
What makes this switch a special device? When the
voltage across the device is turned off--positive or negative--the oxygen bubbles
do not migrate. They will freeze where they have been before, which means that the
boundary between the two titanium dioxide layers is frozen.
That is how the Memristor ``remembers'' the last state of conductivity as well as 
it proves the plasticity properties in memristor to be applied as a synapse in an
artificial neural network architecture and neuromorphic platform.

\subsection{Electrical model}
\begin{figure}[tp]
\centering
\includegraphics [width=7.5cm]{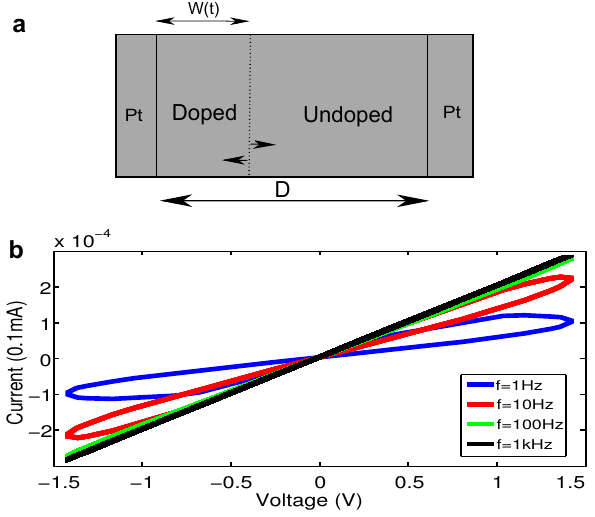}
\caption{\small{Memristor schematic and behavior: \textbf{a)} the memristor structure, 
the difference in applied voltage changes doped and undoped regions,
\textbf{b)} current versus voltage diagram, which demonstrates hysteresis 
characteristic of a memristor, in the simulation we apply the sinusoidal input wave 
with an amplitude of 1.5v,
 different frequencies, $R_{ON}=100\Omega,R_{OFF}=15k\Omega, D=10nm,
 \mu _v=10^{-10}cm^{2}s^{-1}V^{-1}$.}}
 \label{mem_first_fig}
\end{figure}

When an electric field is applied to the terminals of the memristor, the shifting in the boundary between 
its doped and undoped regions leads to variable total resistance 
of the device. In Figure \ref{mem_first_fig}.a, the electrical 
behavior of memristor can be modeled as follows \cite{strukov_missing_2008}:
 \begin{equation} \label{eq:1}
v(t)=R_{mem}i(t)
\end{equation}

\begin{align}
\label{eq:2}
R_{mem}={R_{ON}}\frac{{w(t)}}{D} + {R_{OFF}}(1 - \frac{{w(t)}}{D})
\end{align}
where $w(t)$ is the width of the doped region, $D$ is the overall
thickness of the TiO$_{2}$ bi-layer, $R_{ON}$ is the resistance when
the active region is completely doped ($w=D$) and $R_{OFF}$ 
is the resistance, when the TiO$_{2}$ bi-layer is mostly undopped
($w$$\to$ 0).\\
\begin{equation}
\label{eq:3}
\frac{{dw(t)}}{{dt}}={\mu _v}\frac{{{R_{ON}}}}{D}i(t)
\end{equation}
\text{which yields the following formula for $w(t)$:}

\begin{equation}
\label{eq:4}
w(t)={\mu_v}\frac{{{R_{ON}}}}{D}q(t)
\end{equation}
Where $\mu_v$ is the average dopant mobility. By inserting Equation (\ref{eq:4}) into Equation 
(\ref{eq:2}) and then into Equation (\ref{eq:1}) we obtain the memristance
of the device, which for ${R_{ON}} \ll {R_{OFF}}$ simplifies to:
\begin{equation}\label{eq:5}
R_{mem}=M(q)={R_{OFF}}(1 - \frac{{{\mu _v}{R_{ON}}}}{{{D^2}}}q(t))
\end{equation}
Equation (\ref{eq:5}) shows the dopant drift mobility $\mu{ _v}$
and semiconductor film thicknesses D are two factors with crucial
contributions to the memristance magnitude.
Subsequently, we can write Kirchoff's voltage law for memristor given by:
\begin{equation}
  v(t)=M(q)i(t)
\end{equation}

By using Verilog-A HDL, we simulate the behavior of memristor, 
based on its behavioral equations. 
To investigate the characteristics of memristor in 
electrical circuits, the Verilog-A model of memristor behavior 
must be applied as a circuit element in the HSPICE netlist.
In the HSPICE circuit, we apply a sinusoidal source to 
observe the memristor reaction in a simple circuit consisting 
of the memristor and the sinusoidal source. Figure \ref{mem_first_fig}.b depicts $i-v$ plot 
of memristor terminals that we measured 
in our simulation. This $i-v$ plot, which is the most 
significant feature of memristor \cite{chua-76}, is namely called ``pinched  
hysteresis loop''. The $i-v$ characteristic 
 demonstrates that memristor can ``remember'' the last electric charge 
 flowing through it by changing its memristance. Therefore, we can use the memristor as a latch to
save the data and also as a switch for computing. Moreover, in Figure \ref{mem_first_fig}.b,
it is depicted 
that the pinched hysteresis loop is shrunk by increasing frequency. In fact, 
when the frequency increases toward infinity, memristor behavior is similar to a linear resistor.
\section{Memristor classification based on different materials and applications}
A memristor is generally made from a 
metal-insulator-metal (MIM) sandwich with the insulator usually consisting of a thin film 
like TiO$_{2}$ and a metal electrode like Pt. A memristor can be made from any
Metal Insulator Metal (MIM) sandwich which
exhibits a bipolar switching characteristic.
It means that TiO$_{2}$ and Pt are not the only materials to
fit the criteria for a memristor. For instance, Wan Gee Kim \textit{et al}. 
\cite{Gee-Kim} conducted a systematic approach using the
HfO$_{2}$ and  ZrO$_{2}$ as substitutes for TiO$_{2}$, also  
using TiN or Ti/TiN electrode instead of Pt. Basically, any two-terminal black box is 
called a memristor only if it can present a pinched hysteresis
loop for all bipolar periodic input signals. Following we discuss four
most significant materials for memristor fabrication namely:

 \begin{itemize}
\item Resistive memristor
\item Spintronic memristor
\item Organic (Polymeric) memristor
\item Ferroelectric memristor
 \end{itemize}
 
\subsection{Resistive Memristor}
Before the memristor getting well-known, resistive
materials have already been widely used in the resistive
random access memories (ReRAM/RRAM) \cite{waser_redox-based_2009}. The storage
function of ReRAM is realized by an intrinsic physical
behavior in ReRAM, that is called resistive switching. 
The resistive material 
can be switched between a high resistance state
(HRS) and a low resistance state (LRS) under an external
electrical input signal. 
The TiO$_{2}$ memristor is a ReRAM fabricated in nanometre scale (2-3 nm) thin film that is depicted in Figure 
\ref{mem_first_fig}.a , containing a doped
region and an undoped region. Strukov \textit{et al}. \cite{strukov_missing_2008} exploit a 
very thin-film TiO$_{2}$ sandwiched between two platinum (Pt)
contacts and one side of the TiO$_{2}$ is doped with oxygen vacancies, which are
positively charged ions. Therefore, there is a TiO$_{2}$ junction where one side is
doped and the other is undoped. 
Such a doping process results in two different
resistances: one is a high resistance (undoped) and the other is a low resistance
(doped). The application of an external bias v(t) across the device will move
the boundary between the two regions by causing the charged dopants to drift.
How TiO$_{2}$ could change and store the state has been introduced in \ref{howworks}.

The obvious disadvantage of the first published TiO$_{2}$ memristor was its switching speed
(operate at only 1Hz). The switching speed was not comparable with SRAM, DRAM and even flash memory. 
Flash exhibit writing times of the order of a microsecond and volatile memories have writing speeds of the order 
of hundreds of picoseconds. Many research groups in different labs 
published their fabrication results to demonstrate a faster switching speed device.
In October 2011, HP lab developed a memristor switch using a SET
pulse with a duration of 105 ps and a RESET pulse with a duration of 120 ps. 
The associated energies for ON and OFF switching were
computed to be 1.9 and 5.8 pJ, respectively which are quite efficient for power-aware computations. 
The full-length D (Figure \ref{mem_first_fig}.a) of the TiO$_{2}$ memristor is 10 nm  \cite{HP-speed} that 
proposes high-density devices in a small area in VLSI.

A research team at the University of Michigan led by 
Wei Lu \cite{Wei-syn-mem} demonstrated another type of resistive memristor that can be used to build brain-
like computers and known as amorphous silicon memristor. The Amorphous silicon memristor consists of a layered device
structure including a co-sputtered Ag and Si active layer with a properly designed Ag/Si mixture
ratio gradient that leads to the formation of a Ag-rich (high conductivity) region and a Ag-poor
(low conductivity) region. This demonstration provides the direct experimental support for 
the recently proposed memristor-based neuromorphic systems. 

Amorphous silicon memristor can be 
fabricated with a CMOS compatible simple fabrication process using only common materials
which is a great advantage of using amorphous silicon devices. The endurance test results of
two extreme cases with programming
current levels $10nA$ and $10mA$ are $10^{6}$  and $10^{5}$ cycles respectively. We note the larger than
$10^{6}$ cycles of endurance with low programming currents are already 
comparable to conventional flash memory devices. Wei Lu team have routinely observed switching
speed faster than 5ns from the devices with a few mA on-current.
The switching in this device is faster than 5 ns with a few mA on-current that make it a promising 
candidate for high-speed switching applications.
However, before the devices can be used as a switch, they need to go through a high voltage
forming process (typically $\ge$ 10 V) which significantly reduces the performance of power efficiency 
of devices \cite {si-mem}. Moreover, the retention time (data storage period) is still short (a few months).

\subsection{Spintronic Memristor}
Spintronic memristor changes its resistance by varying the direction of
the spin of the electrons. Magnetic Tunneling Junction (MTJ) has been used in
commercial recording heads to sense magnetic flux. It is the
core device cell for spin torque magnetic random access
memory and has also been proposed for logic devices. In a spintronic device, the electron
spin changes the magnetization state of the device. The magnetization
state of the device is thus dependent upon the cumulative
effects of electron spin excitations \cite{Wang2009Spintronic}. MTJ can be switched 
between a LRS and an HRS using the spin-polarized current
induced between two ferromagnetic layers. If the resistance of this
spintronic device is determined by its magnetization state, we
could have a spintronic memristor with its resistance depending
upon the integral effects of its current profile. 

\begin{figure}
\centering
\includegraphics[width=6cm]{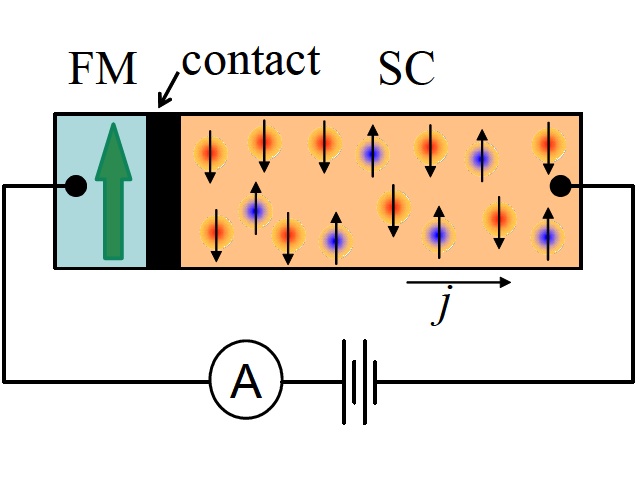}
\caption{Spintronic memristor:Physical schematic of
the circuit made of an interface between a semiconductor and a half-metal 
(ferromagnets with 100\% spin-polarization at the Fermi level) 
(From\cite{spin1}).}\label{fig:Spin}
\end{figure}

The use of a fundamentally
different degree of freedom which allows for the realization of memristive behavior
is thus desirabled by Pershin and Di Ventra \cite{spin1}. They demonstrated that the 
degree of freedom is provided by the 
electron spin and memristive behavior is obtained from the broad class of
semiconductor spintronic devices. This class involves systems whose transport
properties depend on the level of electron spin polarization in a semiconductor which is
influenced by an external control parameter (such as an applied voltage).
Pershin and Di Ventra considered a 
junction with half-metals shown in Figure \ref{fig:Spin} 
(ferromagnets with 100\% spin-polarization at the Fermi level),
because these junctions react as perfect spin-filters, therefore they are more sensitive to the level of
electron spin polarization. They observed memristor behavior in the $i-v$ curve of these
systems. This means the proposed device is controllable
and tunable. Furthermore, the device can be easily integrated on top of the
CMOS. The integration of the spin torque memristor on CMOS
is the same as the integration of magnetic random access memory
cell on CMOS \cite{spin2}. This has been achieved and commercialized in
magnetic random access memory. 

The potential applications of spintronic memristor are 
in multibit data storage and logic, novel sensing scheme, low power consumption computing, 
and information security.
However, the small resistance ON/OFF ratio remains a notable concern for spintronic memristor devices.

\subsection{Organic (Polymeric) Memristor}
\begin{figure}[ht]
\centering
\includegraphics[width=5cm]{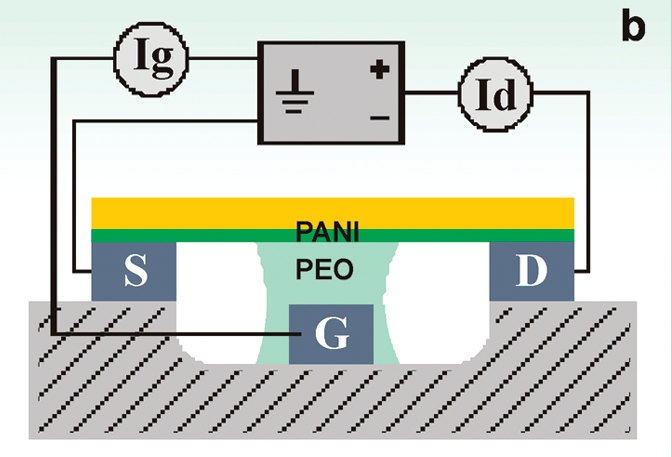}
\caption{Organic (polymeric) Memristor: the active channel is formed by PANI 
on top of a support and two electrodes. The region between PANI and PEO is called the `active zone',
and conductivity transformation is performed here.}\label{fig:poly}
\end{figure}

 In 2005 Erokhin \textit{et al}. \cite{poly2005} at the university of 
Parma reported a polymeric electrochemical element for the adaptive networks.
 Even though it was not called a memristor, however mainly its characteristics  
corresponds to the hypothetical memristor. At the heart of this device, there is a conducting channel, a
thin polyaniline (PANI) layer, deposited onto an insulating
support with two electrodes. A narrow stripe of solid electrolyte
doped Poly Ethylene Oxide (PEO) which is
formed in the central part of the channel and used for the redox reactions. The area of PANI under
PEO is the active zone (see Figure \ref{fig:poly}). A thin silver wire is inserted into the
solid electrolyte to provide the reference potential; such a
wire is connected to one of the electrodes on the solid
support, kept at the ground potential level. 

Conductivity variations and memory properties of the
organic memristor are due to the redox reactions
occurring in the active zone, where PANI is reversibly
transferred from the reduced insulating state into the
oxidized conducting one \cite{Berzina-poly}. In analogy with the nomenclature used in 
Field Effect Transistors (FETs), the two electrodes that are connected with the PANI film 
are called the source and drain electrodes, while the wire immersed in the PEO 
is called the gate electrode.
In the normal operation of the device, the source and the gate electrodes are kept at ground potential, and a
 voltage is applied to the drain electrode. Therefore, we can consider the organic memristor as a two-terminal device.
The polymeric
memristor, compared to the resistive memristor,  can better meet the criteria of the 
theoretical memristor, as its resistance is generally governed
by the charge transfer.

 The Polymeric memristor was investigated in pulse mode, mimicking the synaptic behavior of signal
transmission in neural systems. The phenomenon that the
PANI conductivity, when connected to a positive signal
excitation which is gradually increased similar to the
synapse behavior in real biological neural systems,
described in the Hebbian rule. Simple circuit based on the polymeric
memristor has already been designed to realize both
supervised and unsupervised learning in neural networks \cite{Berzina-poly}.
 Organic materials present several advantages in terms of functionality, the deposition technique, costs
and above all for the relative ease with which the material
properties may be tailored by a chemical approach \cite{organic2}. 
Organic memristor operates with very low power energy. The transformation  
to the conducting state occurs at potentials
 +0.4 - +0.6 V and transformation into the insulating state take place at
potentials lower than +0.1 V \cite{Organic3}.

\begin{figure}[hb]
 \centering
\includegraphics [scale=1.3]{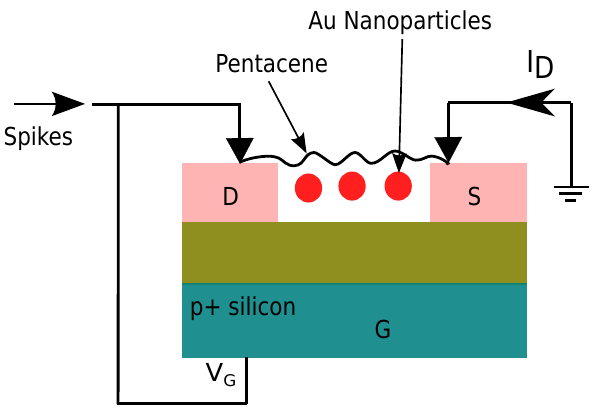}
\caption{Physical structure of the NOMFET. It is composed of a p$^{+}$ doped bottom-gate covered with silicon oxide. 
Source and drain electrodes are made of gold and Au NPs are deposed on the inter-electrode gap before the pentacene
deposition.}
\label{fig:nom}
\end{figure}

There is another type of organic memristor namely Nanoparticle Organic Memory
Field Effect Transistor (NOMFET). NOMFET is made of conjugated molecules and metal nanoparticles (NPs).
This device is initiated and fabricated by the
institute of microelectronics and nanotechnology at Lille university \cite{alibart_organic_2010}. 
NOMFET consists of a bottom-gate and
source-drain contact organic transistor configuration. The gold NPs (5 nm in diameter) were
immobilized into the source-drain
channel by applying self-assembled monolayer covered by a thin film of Pentacene 
as it is shown in Figure \ref{fig:nom}.
The NOMFET has the capability of mimicking synaptic properties as a volatile memory. 
We have used NOMFET in our research to make a 
new type of synapse \cite{shahsavari_combining_2016}. 
Consequently, we will discuss it more in details.

\subsection{Ferroelectric Memristor}
Ferroelectricity is a property of certain materials which have a spontaneous electric polarization that can be
reversed by the application of an external electric field.
The ferroelectric memristor is based on a thin ferroelectric barrier sandwiched between two metallic electrodes. 
Therefore, these two opposite polarization states can be used to represent
binary bits `0' and `1', thus resulting in the advent of the
Ferroelectric Random Access Memory (FeRAM). Due to
FeRAM nonvolatility, ferroelectric materials have
been widely used in automobile equipment, ID/smart card, Radio Frequency Identification
(RFID) and other embedded memory applications \cite{wang_next_2014}.

\begin{figure}[h]
\centering
\includegraphics[scale=1]{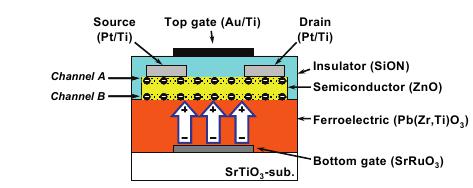}
\caption{{\small Ferroelectric Memristor, the OxiM transistor has dual channels at the upper and
lower sides of the ZnO film, which are controlled independently by the top gate and the
bottom gate, respectively. The bottom FET has the gate (SRO layer) 
and insulator (PZT ferroelectric layer) constituting a FeFET that has
memory characteristics (from \cite{Ferro2}). }}\label{fig:ferro}
\end{figure}

Chanthbouala, A. \textit{et al}. \cite{ferro}, showed voltage-controlled
domain configurations in ferroelectric tunnel barriers yield
memristive behavior with resistance variations exceeding two 
orders of magnitude and a 10 ns operation speed.
They reported Piezoresponse Force Microscopy (PFM) images and
electrical transport measurements as a function of the amplitude,
duration and the repetition number of voltage pulses in the 10-200 ns
range.
In tunnel junctions with a ferroelectric barrier, switching the 
ferroelectric polarization induces variations of the tunnel resistance,
with resistance contrasts between the ON and OFF states of several
orders of magnitude. The low-resistance state ($R_{ON}$) corresponds to
the ferroelectric polarization pointing up ($P\uparrow$), and the 
high-resistance state ($R_{off}$) corresponds to
the ferroelectric polarization pointing down ($P\downarrow$). 
% The nanodevices with diameters
% of 350nm were defined from BTO/LSMO (BaTiO$_{3}$/La$_{0.67}$Sr$_{0.33}$MnO$_{3}$)
%  bilayers by electron-beam lithography and
% lift-off of sputter-deposited Co (10 nm) followed by a capping layer of Au (10 nm). 
Positive and negative trains of pulses applied consecutively with +2.9 V and -2.7 V amplitude. 
In the analogy with the
operation of FeRAM, the large
ON/OFF ratio in ferroelectric tunnel junctions (FTJs) has so far
been considered only for binary data storage, with the key advantage
of non-destructive readout and simpler device architecture, however still non-CMOS compatible.
 In another study, Y.Kaneko \textit{et al}.\cite{Ferro2} presented a new transistor and 
 implemented it by all oxide-based ferroelectric thin films,
 which include SrRuO$_{3}$ (SRO: bottom gate electrode), Pb(Zr,Ti)O$_{3}$ (PZT: ferroelectric), 
 ZnO (semiconductor), and SiON (gate insulator) Figure \ref{fig:ferro}. They have demonstrated the 
 conductivity modulation of the
interface between two oxides, ZnO and PZT, in a FeFET, is
applicable for a nonvolatile memory which has the same memristive operation.
Ferroelectric-based memristor cell can be expected
to be very suitable for the nonvolatile memory array configuration
and the future neuromorphic systems embedded in
the intelligent transparent electronic applications \cite{ferro3}. 

\subsection{Evaluation of Memristor with different materials}

\begin{figure}

\centering

\begin{tikzpicture}
\begin{axis}[
    symbolic x coords={Spintronic, Ferro, Resistive, Organic },
    bar width=1.4 cm,
        ylabel = {Number of scientific publications},
        xlabel = {Different types of memristor devices},
    xtick=data,enlarge x limits]
    \addplot[ybar,fill=blue] coordinates {
        (Spintronic,38)
        (Ferro,42)
        (Resistive,186)	
    (Organic,72)
       };
\end{axis}
\end{tikzpicture}
\caption{Number of publications for each type of memristors.}
\label{fig:8}
\end{figure}
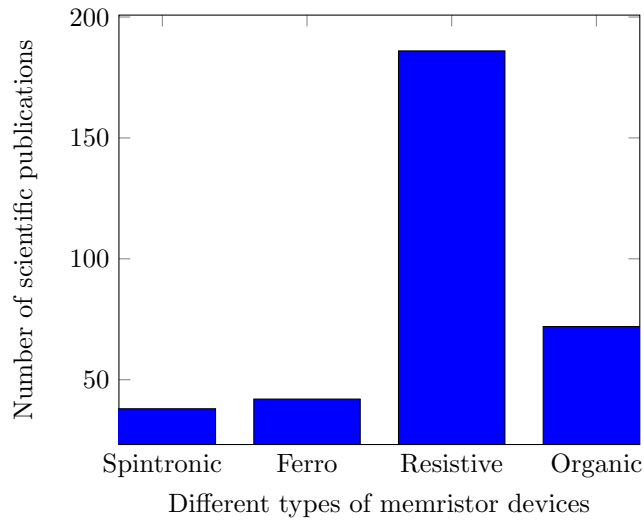

After 2008, there have been many articles related to memristors and memristive devices.
Here we evaluate the four types of memristors  
using different materials. TiO$_{2}$ memristor  is the first fabrication of memristor 
device which is considered as one of the most promising ones.
There are hundreds of publications related to TiO$_{2}$ memristor .
Recently, Stanley Williams research group
enhanced the characteristics of TiO$_{2}$ memristor such as switching speed, programming endurance, 
and retention time which made TiO$_{2}$ memristor the best candidate to apply in commercial usages. 
Chen Yirin \cite{Wang2009Spintronic, spin2} assistant professor at the University of Pittsburgh, 
published 49 articles
in memristor technology and application are working on spintronic memristor,
Pershin and Di Ventra \cite{spin1} are the other researchers who trying to
improve spintronic memristive devices characteristics to use it as 
a nonvolatile memory. Organic materials present several 
advantages, therefore, could be the next proper candidate to fabricate memristive devices. 
The last material that we have evaluated here is ferroelectric memristor. 
After publishing a ferroelectric memristor in 
Nature materials journal \cite{ferro} this type of 
fabrication is introduced as another option for building Memristor. 
We sum up specifications and characteristics of these five classes 
of memristor material in Table \ref{tab:1}. 
Lei Wang \emph{et al.} \cite{wang_overview_2015} discussed 
another type of memristors such as manganite memristor and 
Resonant-Tunneling Diode (RTD) memristor. We note that 
the magnetic memristor has the similar behavior to the
ferroelectric type of memristive devices. RTD has more
potential to react as a complementary 
device beside memristor for neural network applications e.g., 
Cellular Neural Networks (CNN) \cite{hu_multilayer_2015}.
A quantitative comparison respecting the number of published 
paper specifically in fabrication and material of different 
classes of the memristors is presented in Figure \ref{fig:8}. 
   
\begin{table}[hb]

\begin{center}
%\rowcolors{1}{brown}{gray}
\scalebox{0.7}{
\begin{tabular}{|p{2cm}|p{4cm}|p{3cm}|p{3cm}|p{3.5cm}|}

\hline
Memristor&Advantage&Disadvantage&Applications&University-Lab\\
\hline
Resistive & small scale, fast switching, simple structure&still non-reliable for
commercial&memory, logic gates, neuromorphic, analog devices &HP Lab\\
\hline
Spintronic&magnetic memory match technology &the small
resistance ON/OFF ratio& Neuro-inspired systems, memory
&University of Pittsburgh, US\\ 
\hline
Organic&relative ease with chemical materials, 
work with ultra-low power&slow switching&artificial synapse&Parma/Lille University\\
\hline
Ferroelectric&suitable for the nonvolatile memory array&slow switching, 
Non-CMOS Compatible& synapse, RRAM&Panasonic, Japan \& Thales, France\\
\hline

\end{tabular}
}
\end{center}

\caption{Table of different class of memristors based on different materials and its applications, the first 
university/lab announcement of the device is listed too.}
\label{tab:1}
\end{table}

\label{materialClass}

\section{Potential applications of memristors}
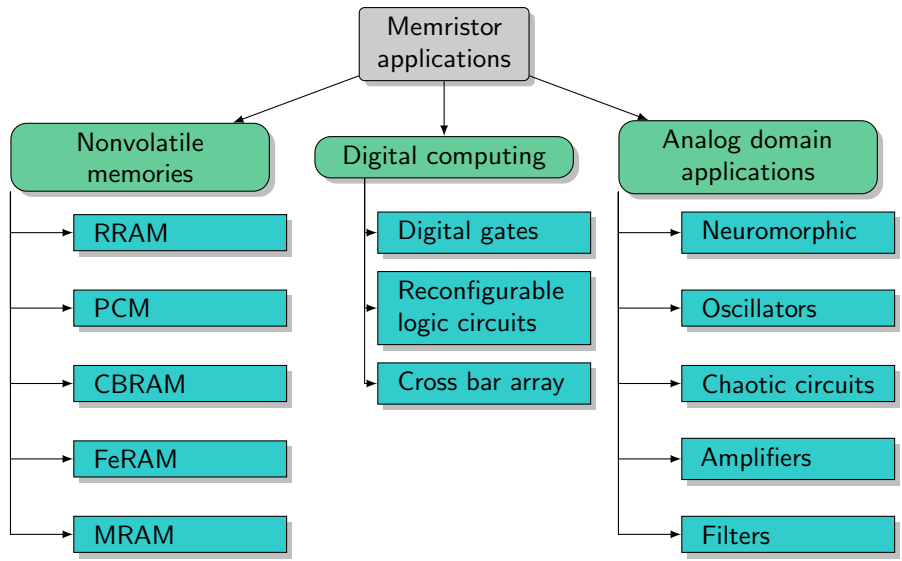
\begin{figure}[hb]
\centering
\begin{tikzpicture}[
  level 1/.style={sibling distance=40mm,minimum width=3.4 cm},
  edge from parent/.style={->,draw},
  >=latex]

% root of the the initial tree, level 1
\node[root] {Memristor applications}
% The first level, as children of the initial tree
  child {node[level 2] (c1) {Nonvolatile memories}}
  child {node[level 2] (c2) {Digital computing}}
  child {node[level 2] (c3) {Analog domain applications}};

% The second level, relatively positioned nodes
\begin{scope}[every node/.style={level 3},minimum width=2.8cm]
\node [below of = c1, xshift=15pt] (c11) {RRAM};
\node [below of = c11] (c12) {PCM};
\node [below of = c12] (c13) {CBRAM};
\node [below of = c13] (c14) {FeRAM};
\node [below of = c14] (c15) {MRAM};

\node [below of = c2, xshift=15pt] (c21) {Digital gates};
\node [below of = c21] (c22) {Reconfigurable logic circuits};
\node [below of = c22] (c23) {Cross bar array};
%\node [below of = c23] (c24) {Using overlays};

\node [below of = c3, xshift=15pt] (c31) {Neuromorphic};
\node [below of = c31] (c32) {Oscillators};
\node [below of = c32] (c33) {Chaotic circuits};
\node [below of = c33] (c34) {Amplifiers};
\node [below of = c34] (c35) {Filters};
\end{scope}

% lines from each level 1 node to every one of its "children"
\foreach \value in {1,2,3,4,5}
  \draw[->] (c1.195) |- (c1\value.west);

\foreach \value in {1,...,3}
  \draw[->] (c2.195) |- (c2\value.west);

\foreach \value in {1,...,5}
  \draw[->] (c3.195) |- (c3\value.west);
\end{tikzpicture}
\label{diagram:1}
\caption{Different memristor applications in different domains.}
\label{fig:1-9}
\end{figure}

In the previous section, we have discussed the different type of memristive materials. In this section, we study the
potential application of memristors. 
we divided the applications into three main classes:
nonvolatile memories, digital computing, and analog domain applications. The classification in addition to 
the practical application examples is depicted in Figure \ref{fig:1-9}.
This classification may cover most of the recent applications, however, the memristor is a novel device which new 
capabilities may be introduced soon in 
various areas of research. Therefore, the novel applications are dramatically anticipated. 

%Diagram

\subsection{Memristor-based nonvolatile memory}
Memristor-based nonvolatile memory is the most obvious application of memristors \cite{Hamdioui1_mem}. 
The nonvolatile, memristor-based memory cell compared to SRAM and DRAM, can exhibit non-volatility, 
good scalability, compatible with
conventional CMOS electronics, and the last but not the least, it has no leakage power.
Several types of memories are introduced using memristor nanodevice. Resistive RAM (RRAM) \cite{RRAM}, Phase 
change memory (PCM) \cite{driscoll_phase-transition_2009}, 
Conductive Bridge memory (CBRAM) \cite{liu_controllable_2010}, ferroelectric memory (FeRAM) \cite{ferro}
and the spintronic memristor
that can be a promising replacement for 
Magnetic memory (MRAM) \cite{MRAM}. 
A memory array of Memristors what is called a resistive RAM or RRAM is another form of memristor memory.
 RRAM operates faster than phase-change memory (PCRAM), and it has simpler and smaller cell structure
 than magnetic memory (MRAM).   
 
\subsection{Digital computing}
Another possible application of memristor is digital design and computation. Memristors can be applied 
in hybrid CMOS-memristor circuits, or as a basic component to built the logic gates \cite{Logic-Wiliams}.
One remarkable logic
application is using memristor as a reconfigurable switch (FPGA) \cite{FPGA}. The implication logic 
synthesis \cite{LehtonenPL12} and crossbar array architecture \cite{Crossbar-mem} are two 
alternatives presented a the new approaches to make efficient digital gate library.
A comprehensive review and implementation of both digital computing approach is 
discussed in \cite{shahsavari_unconventional_2015}.
\subsection{Analog domain applications} 
Another further research area of memristive devices is
analog domain application. Simple circuits of memristors
with a single capacitor or inductor are discussed in \cite{analog1}. 
There are several research work in analog domain using memristors
such as neuromorphic computing, amplifiers, 
memristor oscillators, filters and chaotic circuits  (see
e.g. \cite{Amplifier}, \cite{Oscillator}, \cite{analogICs}, \cite{chaotic}). 

Most interesting and recent study field in analog domain that we will 
focus on it more in our research is neuromorphic or neuro-inspired computation. It
is believed this approach is promising for yielding
innovations in future cognitive computing, hence will get probably more attention
in the  near future of research.
The key to the high efficiency of biological neural systems 
is the large connectivity between neurons that
offers highly parallel processing power. 
Another key factor of high efficiency of biological neural network is 
the connection plasticity.
The synaptic weight between two neurons can be precisely
adjusted by the ionic flow through them and it is widely
believed that the adaptation of synaptic
weights enables the biological neural systems to learn. 
Memristors function similarly to the biological synapses. 
This characteristic makes
Memristors proper building blocks in neuromorphic systems, where neurons and
synapses are modeled as electronic devices \cite{Wei-syn-mem}. 
Memristors can be made extremely small, therefore, 
by applying them, mimicking the functions of a brain
would be possible in near future \cite{Williams-missingmem}.  
The brain-like computing capability of the
memristor-based complex spike timing-
dependent plasticity (STDP)
learning networks is demonstrated 
by Afifi \textit{et al} \cite{Afifi-stdp}.
Memristor could be applied in more efficient learning platform such as
Deep Belief Networks for online learning \cite{neftci_event-driven_2014,fatahi_towards_2016}.
The model of memristor is used in neuromorphic hardware simulation e.g.,
Xnet \cite{bichler_design_2013} and N2S3 \cite{shahsavari_n2s3_2015}.
\begin{figure}[b]
\centering
\includegraphics[scale=.4]{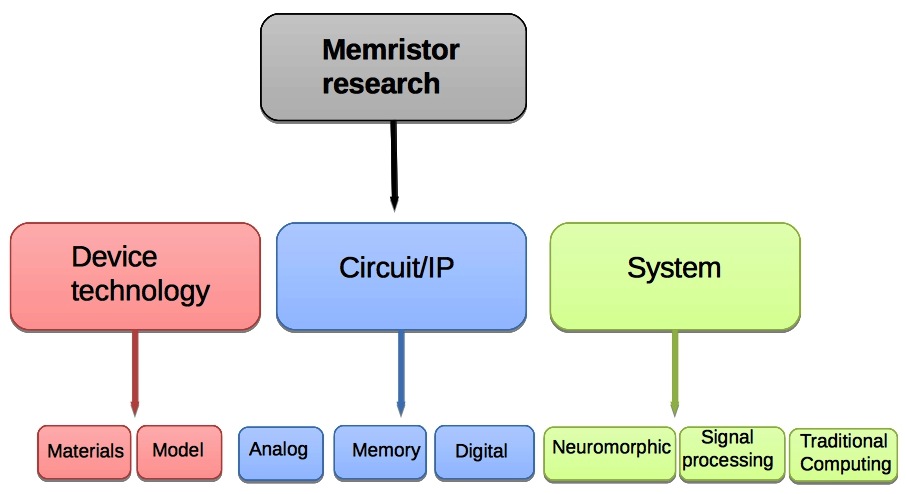}
\caption{Classification of domain studies using memristive devices}\label{chart}
\end{figure}
\section{Streams of research}
The ability to indefinitely store resistance values means that a memristor 
can be used as a nonvolatile memory.
There are yet more potential applications that we did not mention previously
, we point to the capacity of memristive 
networks in realizing demanding image processing
and more specifically edge detection and face/object recognition \cite{Image}. 
Pershin and Di Ventra address the capability of memristors
to perform quantum computation
in addition to conventional neuromorphic
and digital data processing \cite{Quantum}. 
Performing arithmetic operations in memristor-based 
structures is possible in both analog and digital approaches \cite{Arith}. 
Another system level application is memristive Neuro-Fuzzy System \cite{Fuzzy}.
However, the memristor potential goes 
far beyond instant-on computers to embrace one of 
the biggest technology challenges:
mimicking the functionality of the brain. Within a decade, 
memristors could let us emulate,
instead of merely simulate, networks of 
neurons and synapses. By replacing several specific transistors with a 
crossbar of memristors, 
circuit could be shrunk by nearly a 
factor of 10 in area and improved in terms of its speed relative to 
power consumption 
performance \cite{Williams-missingmem}. 
Computing with memristor is another interesting approach which 
memristor plays a crossbar switch role in the circuit. Using memristor-based 
circuit for performing arithmetic operations, signal processing application, 
dynamic load, oscillators, amplifiers, sensing application, 
artificial biological system, image encryption, 
and many other approaches are several  
applications of memristor that recently appeared in 
different research publications. 
In this context, we made a simple classification of a stream of 
research as shown in Figure \ref{chart} 
that could be useful for those who want to use this 
flexible device in their research studies. 
The hybrid approaches are not in the chart, 
here we mention that any sort of
utilization can be implemented in hybrid circuit
instead of using pure memristive circuit. 
\section{Conclusions and summary}
In this study, we have done a feasibility study on memristive nanodevice 
from theoretical model to practical applications. 
we studied the cons and pros of using different memristor devices based on 
the applied material to manufacture the device. As each 
material or type of memristor has its own characteristics,
consequently one can analyze and 
discover the different potential of application of each type of
these memristive devices more conveniently. The better understanding of 
physics of different memristive materials leads to discover 
more sophisticated applications of these devices.
Four general memristive devices are surveyed namely:
Resistive, Spintronic, Organic (polymeric) and Ferroelectric memristive devices.
The potential application as well as advantages versus disadvantages 
of using each one are presented too.
The resistive memristor has been applied more than others
in different research works from memory to artificial synapse. 
The Spintronic and Ferroelectric devices show 
promising properties to make new nonvolatile memories.
The organic memristor is more appropriate to make 
artificial synapse in Spiking Neural Networks.
The mix combination of those materials not only
propose new research studies but also take the 
advantages of using more useful properties of each device.

The practical applications of memristive devices are presented
subsequently. Three main domains of potential applications have been classified:
a)nonvoaltile memory such as RRAM, PCM, CBRAM, FeRAM and MRAM;
b)digital computing domain such as logic implication and crossbar array;
c)analog domain of application such as neuro-inspired computing in 
spiking neural networks, oscillators, filters and amplifiers.
Among these applications, we focus on computation approaches
by using two digital and neuromorphic domains.
\bibliographystyle{unsrt}
%\bibliography{unsrt}
\bibliography{mythesis}
\end{document}